# Time Series Analysis of the Southern Oscillation Index using Bayesian Additive Regression Trees

*S. van der Merwe, Department of Mathematical Statistics and Actuarial Science, IB75, University of the Free State, Box 339, Bloemfontein, 9300, South Africa*
*October 2009*

**Abstract**

Bayesian additive regression trees (BART) is a new regression technique developed by Chipman *et al.* (2008). Its usefulness in standard regression settings has been clearly demonstrated, but it has not been applied to time series analysis as yet. We discuss the difficulties in applying this technique to time series analysis and demonstrate its superior predictive capabilities in the case of a well know time series: the Southern Oscillation Index.

## Introduction

The Southern Oscillation Index (SOI) measures the difference in air pressure between Tahiti and Darwin. This value has a strong influence on weather patterns. In De Waal (2009) we see how the October SOI value in particular affects the rainfall and river flows in South Africa. Being able to accurately predict this value into the future is thus very important.

In the past the value of the SOI has been predicted using a classical ARMA approach (Chu & Katz, 1985). This approach is limited by the fact that is a linear model of the past values and errors. Non-linear models, on the other hand, present unique challenges to the researcher wishing to apply them. A good example is the use of Neural Networks for time series analysis (Giordano, *et al.* 2007).

The biggest challenge is creating prediction intervals, as the standard method does not extend to complex non-linear models. Alternatives exist, but they are computationally expensive. BART models do naturally produce a predictive posterior distribution for predictions, but extending these predictive intervals further than one time interval into the future is not a simple matter.

According to Chipman *et al*. (2008), BART models work by summing a large number of small decision trees where each tree explains a small portion of the variation in the target variable. These trees are kept small by placing a strong prior distribution on the size of each tree. The trees themselves differ from standard decision trees in that they are built randomly using a MCMC back-fitting technique. They go on to show that these models have extraordinary predictive power when compared to established techniques.

In order to predict the SOI it may be useful to gain a deeper understanding of the properties of the series. It seems to be a naturally oscillating series with multiple long term cycles, as seen by the extreme peaks in the periodogram below. It is worth investigating the effect of including these cycles in any model one may fit to this series.

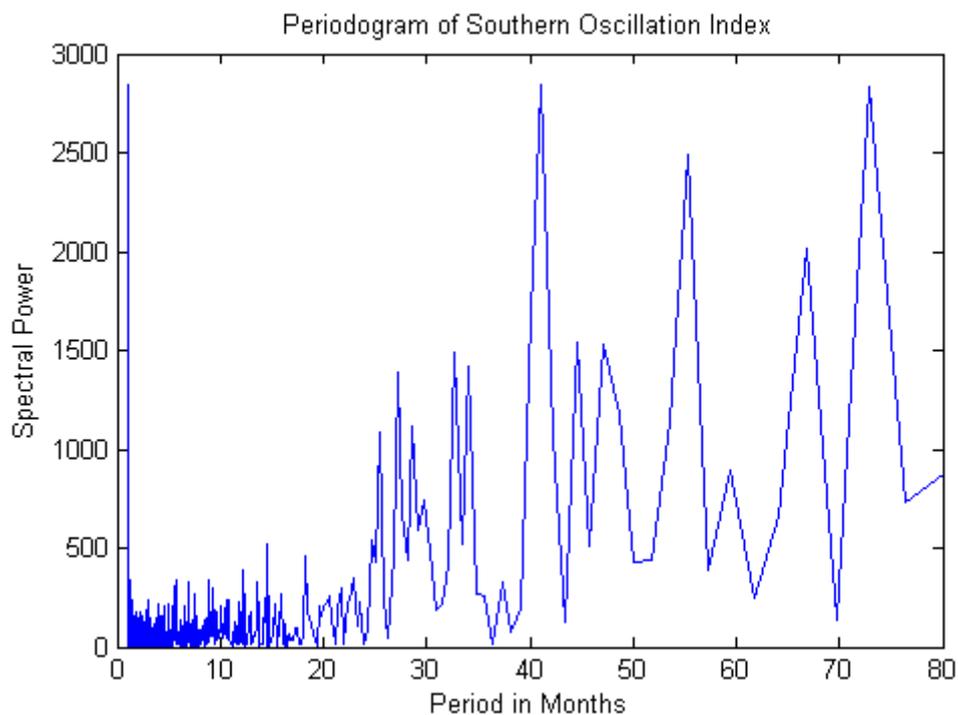

The values of the SOI that were used in this analysis were the values from January 1876 until August 2009, making this series fairly long by time series standards. However, BART is by its very nature a data mining model that works best when presented with very large samples. It is this nature that also makes these models prone to over-fitting. We thus treat the modelling of the series using BART as a data mining problem but randomly dividing the sample into a training portion and a testing portion.

# Regression Approach to Predicting October's SOI Value

## Calculation of Model Accuracy

Various measures of model accuracy were calculated by comparing the predictions with the observed values for all past Octobers, but distinguishing between those values that were used to fit the model (sample values) and those values that were kept aside (out-of-sample values). Three of these are reported here for every model: the correlation coefficient (CORR), the mean absolute error (MAE) and the root mean square error (RMSE).

It is important to note that BART models are random models, in that they produce different results every time they are fitted. Also, the allocation of observations to the sample versus out-of-sample is random. Thus, all results and measures reported are an average taken over ten fits, or runs, of the model.

## Using all available variables

If we build a model to predict October's value using the twelve months prior as well the as 41st and 73rd months prior to each October we notice the presence of over-fitting, that is to say that the out-of-sample predictions are significantly weaker than the sample predictions.

The model mentioned above with 14 explanatory variables produces the following mean fit statistics:

| Sample | CORR | MAE | RMSE |
|---|---|---|---|
| Training | 0.92 | 3.23 | 4.04 |
| Testing | 0.72 | 4.89 | 6.18 |

Note that for all models in this section a random 80% of the observations are used to train the model and the remaining 20% are used for testing.

## Variable Selection

One of the reasons BART models over-fit is because too many trees are used, but that is not the case here, as we have reduced the number of trees to just twenty. In fact, the number of trees used does not seem to have much impact on the results at all. The results are only slightly weaker when say ten or forty trees are used.

The reason for overfitting here seems to be that too many variables are included in the model. We need to perform some form of variable selection, but we need to bear in mind that we are working with a heavily non-linear model and that most standard variable selection techniques are not appropriate. Thankfully, BART has its own, unique, method of variable selection.

It works by reducing the number of trees to the point where the model has to be selective about which variables to use in each branch. The model is then forced to seek out those variables that are most useful in predicting the target variable. Counting the number of times each variable is used in the model produces a measure of the relative importance of each variable.

Applying this technique to our model produces the following relative variable importance (1 = medium relative importance):

| Variable   | Sep  | Aug  | Jul   | Jun   | May   | Apr  | Mar  |
|------------|------|------|-------|-------|-------|------|------|
| Importance | 3.04 | 1.21 | 1.68  | 1.35  | 1.00  | 0.60 | 0.49 |
| Variable   | Feb  | Jan  | Dec-1 | Nov-1 | Oct-1 | -41  | -73  |
| Importance | 0.67 | 0.42 | 0.96  | 0.58  | 0.44  | 0.93 | 0.62 |

## Reduced Model

Based on this we thus reduce our choice of variables to only the five months prior to each October. This reduces the overfitting problem and gives results as follows:

| Sample   | CORR | MAE  | RMSE |
|----------|------|------|------|
| Training | 0.90 | 3.45 | 4.33 |
| Testing  | 0.81 | 4.89 | 5.90 |

Note that these results can be improved by fitting the model multiple times and selecting the best model.

There is, however, one glaring problem with this model: we only gain one month by using it. We must have September's value in order to predict October's value.

## Excluding September

We can gain another month by excluding September for the previous model. This model is slightly weaker but is, nevertheless, the best model we can produce under these constraints for the purpose of predicting the October value. Its fit statistics are as follows:

| Sample   | CORR | MAE  | RMSE |
|----------|------|------|------|
| Training | 0.82 | 4.49 | 5.60 |
| Testing  | 0.66 | 5.65 | 7.16 |

## Predictions from further back in time

Making predictions using only higher lags is very difficult. We have already seen that there is some relationship with the previous December as well as May of three years prior but we were unable to find additional useful variables. Thus, we can, at best, achieve results as follows:

| Sample | CORR | MAE | RMSE |
|--------|------|------|------|
| Training | 0.49 | 7.10 | 8.58 |
| Testing | 0.22 | 7.80 | 9.40 |

It is possible to achieve better predictions for October by not focusing on October itself. This is especially true in the first half of each year.

## Autoregressive Time Series Approach

In this section we consider every month equal and try to make predictions for the months ahead, completely ignorant of the current date. This is closer to the classic univariate time series approach.

When we use this approach we have twelve times as many observations and so over-fitting is less of a concern (but still worth baring in mind). Here we increase the number of trees in each model to 40. We also keep aside a random third of the observations for testing (as opposed to a fifth).

Again, we need to determine the current variables to include in the model. In time series analysis, one method that is often used is the inspection of the correlograms. If we look at the correlograms below and we restrict ourselves to autoregressive models then it is clear that an AR(5) model is appropriate.

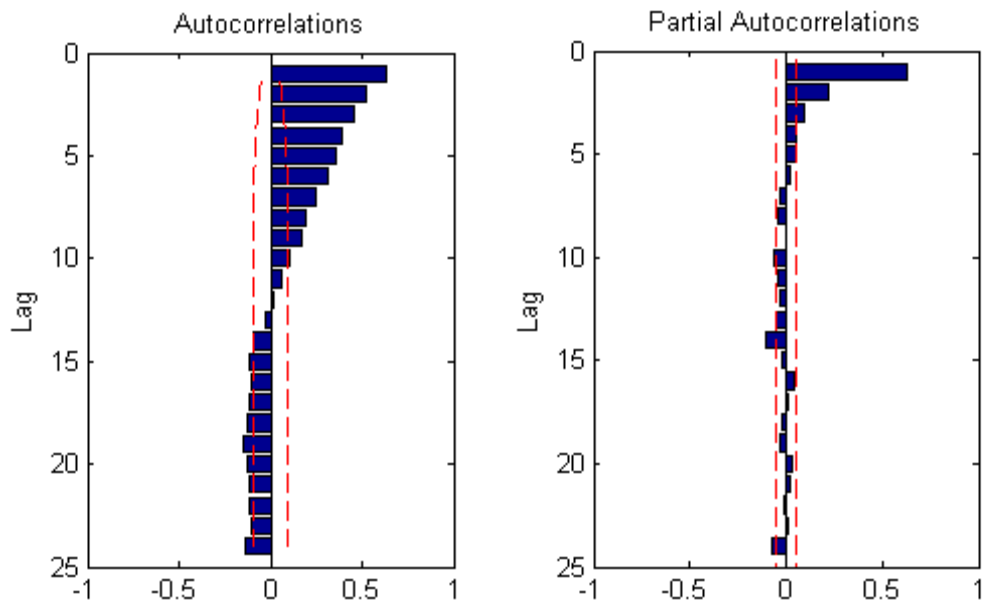

However, autocorrelations are a linear measure of dependence, and so we once again apply BART variable selection:

| Lag | 1 | 2 | 3 | 4 | 5 | 6 |
|---|---|---|---|---|---|---|
| Importance | 4.71 | 2.11 | 1.22 | 0.47 | 0.83 | 0.39 |
| Lag | 7 | 8 | 9 | 41 | 73 | |
| Importance | 0.19 | 0.29 | 0.21 | 0.42 | 0.17 | |

From the table it is clear that using the first 5 lags is, in fact, appropriate. Making predictions into the future using BART models of this form gives the following out-of-sample fit statistics:

| Out of Sample | Correlation | MAE | RMSE |
|---|---|---|---|
| **1 Month** | 0.647 | 6.158 | 7.886 |
| **2 Months** | 0.552 | 6.763 | 8.624 |
| **3 Months** | 0.483 | 7.083 | 9.069 |
| **4 Months** | 0.425 | 7.365 | 9.381 |
| **5 Months** | 0.385 | 7.554 | 9.555 |
| **6 Months** | 0.346 | 7.697 | 9.710 |
| **7 Months** | 0.304 | 7.852 | 9.880 |
| **8 Months** | 0.253 | 7.980 | 10.041 |
| **9 Months** | 0.239 | 8.001 | 10.076 |
| **10 Months** | 0.220 | 8.052 | 10.122 |
| **11 Months** | 0.233 | 7.967 | 10.069 |
| **12 Months** | 0.218 | 8.014 | 10.121 |

In the table above the predictions are made using the mean of the posterior distribution.

If we use the median of the predictive posterior instead then we get worse results:

| Out of Sample | Correlation | MAE | RMSE |
| --- | --- | --- | --- |
| 1 Month | 0.638 | 6.238 | 7.978 |
| 2 Months | 0.542 | 6.849 | 8.703 |
| 3 Months | 0.477 | 7.130 | 9.114 |
| 4 Months | 0.420 | 7.420 | 9.417 |
| 5 Months | 0.374 | 7.651 | 9.633 |
| 6 Months | 0.325 | 7.779 | 9.825 |
| 7 Months | 0.281 | 7.918 | 9.975 |
| 8 Months | 0.243 | 8.024 | 10.094 |
| 9 Months | 0.207 | 8.110 | 10.193 |
| 10 Months | 0.203 | 8.084 | 10.171 |
| 11 Months | 0.198 | 8.093 | 10.197 |
| 12 Months | 0.193 | 8.093 | 10.202 |

Making predictions into the future with BART models is a great deal more difficult than it is with standard time series models as BART models do not currently allow for situations where the output is to be fed back through the model as input.

If we are merely interested in a point estimate (as above) then we can take the mean prediction as fact and re-fit the model with this value added onto the end of the series. This produces a prediction for two months ahead of the current month, which can, in turn, be added onto the series, *etc*.

If we require predictive intervals then the above process must be repeated many times using a random value from the predictive posterior distribution instead of the mean or median prediction. This process requires a distributed computing environment to avoid excessive run times on current generation computing hardware.

It is worth noting that this model produces white noise residuals when making one-month-ahead predictions. This can be seen quite clearly in the correlogram below:

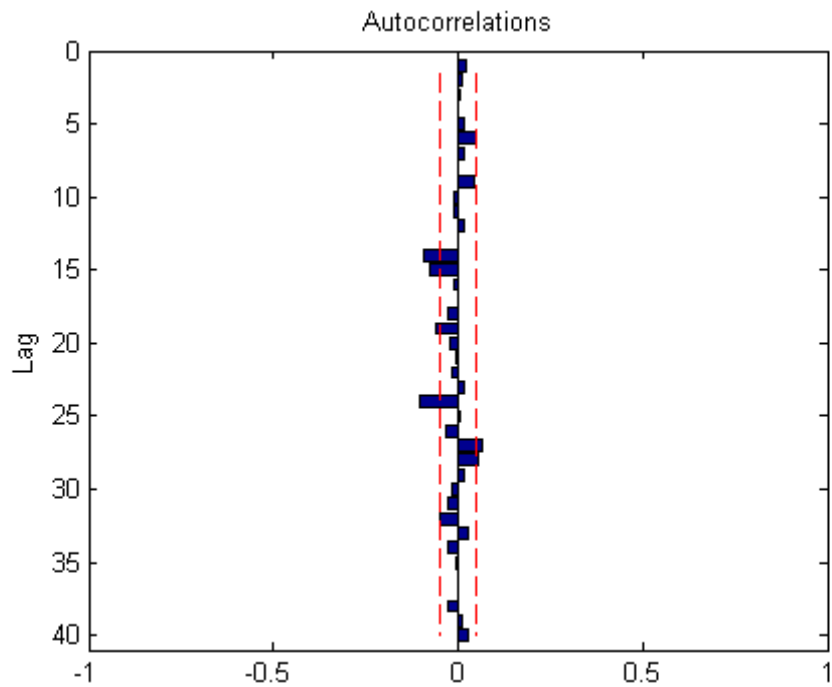

## Comparison and Conclusion

If we fit a linear AR5 model in the same way we obtain the following out-of-sample fit statistics:

| Out of Sample | Correlation | MAE | RMSE |
|---|---|---|---|
| 1 Month | 0.620 | 6.154 | 7.937 |
| 2 Months | 0.532 | 6.743 | 8.564 |
| 3 Months | 0.422 | 7.207 | 9.197 |
| 4 Months | 0.374 | 7.383 | 9.402 |
| 5 Months | 0.338 | 7.466 | 9.524 |
| 6 Months | 0.261 | 7.676 | 9.815 |
| 7 Months | 0.171 | 7.879 | 10.090 |
| 8 Months | 0.156 | 7.807 | 10.059 |
| 9 Months | 0.163 | 7.705 | 9.991 |
| 10 Months | 0.164 | 7.745 | 9.980 |
| 11 Months | 0.148 | 7.761 | 10.016 |
| 12 Months | 0.112 | 7.784 | 10.080 |

This is worse than the BART model by some way. However, if we look at the BART predictions for the coming months we encounter a problem:

| Sep 09 | Oct 09 | Nov 09 | Dec 09 | Jan 10 | Feb 10 |
|--------|--------|--------|--------|--------|--------|
| -1.97  | 0.40   | 0.77   | 0.09   | 1.15   | 0.71   |
| Mar 10 | Apr 10 | May 10 | Jun 10 | Jul 10 | Aug 10 |
| 0.76   | 0.66   | 0.61   | 0.75   | 0.73   | 0.67   |

It appears as though the model gradually tends towards a flat model with the variance fading after 5 months. This may produce a relatively good fit in statistical terms but it is of little value to the researcher wishing to know what future values of the series are likely to be.

We obtain similar (but worse) results from the AR(5) model:

| Sep 09 | Oct 09 | Nov 09 | Dec 09 | Jan 10 | Feb 10 |
|--------|--------|--------|--------|--------|--------|
| 0.27   | 0.55   | 0.06   | 0.04   | 0.15   | 0.10   |
| Mar 10 | Apr 10 | May 10 | Jun 10 | Jul 10 | Aug 10 |
| 0.09   | 0.07   | 0.05   | 0.04   | 0.03   | 0.02   |

In conclusion, the SOI remains a very difficult time series to predict, in spite of the power of BART models.